\documentclass[useAMS,usenatbib,twocolumn]{mn2e}
\usepackage[dvips]{graphicx}

\title[X-ray clusters of galaxies in conformal gravity]{X-ray clusters of galaxies in conformal gravity}

\author[Diaferio \& Ostorero]{Antonaldo Diaferio$^{1,2}$\thanks{E-mail: diaferio@ph.unito.it (AD); ostorero@ph.unito.it (LO)} and Luisa Ostorero$^{1,2,\star}$\\
$^1$Dipartimento di Fisica Generale ``Amedeo Avogadro'', Universit\`a degli Studi di Torino, Via P. Giuria 1, I-10125, Torino, Italy\\
$^2$Istituto Nazionale di Fisica Nucleare (INFN), Sezione di Torino, Via P. Giuria 1, I-10125, Torino, Italy\\
}

\begin{document}
\maketitle

\begin{abstract}
We run adiabatic $N$-body/hydrodynamical simulations of
isolated self-gravitating gas clouds to test whether 
conformal gravity, an alternative theory to General Relativity, is able to explain
the properties of X-ray galaxy clusters without resorting to dark matter.
We show that the gas clouds rapidly reach equilibrium 
with a density profile which is well fit by a $\beta$-model whose
normalization and slope are in approximate agreement
with observations. However, conformal gravity fails to yield the
observed thermal properties of the gas cloud: (i) the mean temperature
is at least an order of magnitude larger than observed; (ii) the temperature
profiles increase with the square of the distance from the cluster
center, in clear disagreement with real X-ray clusters. 
These results depend on a gravitational potential whose
parameters reproduce the velocity rotation curves of spiral galaxies. 
However, this parametrization stands
on an arbitrarily chosen conformal factor. It remains to be seen
whether a different conformal factor,
specified by a spontaneous breaking of the conformal symmetry, can reconcile this
theory with observations.  
\end{abstract}


\section{Introduction}
\label{sec:intro}

On the scale of individual galaxies and larger scales,
General Relativity requires large amounts of
dark matter to describe the  
dynamics of cosmic structure. Moreover, the late-time 
acceleration of the Hubble expansion implies the existence
of a cosmological constant, a special case of
a dark energy fluid which is suggested by more sophisticated models 
(see \citealt{copeland06}, for a review).
In principle, we can avoid the dark matter and dark energy solutions 
to the puzzles posed by the astrophysical data by adopting 
an alternative theory of gravity, which
reduces to General Relativity in the appropriate limit. 
Independently of the dark matter and dark energy problems, a modification
of General Relativity 
is also highly desirable if we ultimately wish to unify gravity with
the other fundamental interactions. 

Possible modifications of the Einstein-Hilbert action proposed
in the literature are, among others, (i) the introduction of additional 
scalar and/or vector fields (e.g., \citealt{fujii03, beken06}); (ii) 
the assumption of arbitrary functions $f(R)$ of the Ricci scalar $R$ (e.g.,
\citealt{capozz-franc07, nojiri08});
(iii) the introduction of additional dimensions to the four dimensions of the
General Relativity spacetime manifold (e.g., \citealt{maart04}).

A different approach was suggested by \citet{mannh90} who
revived  Weyl's theory \citep{weyl18, weyl19, weyl20} as a possible candidate to solve
the dark matter and dark energy problems. When the geometry is
kept Riemannian, with a null covariant derivative of the metric tensor, 
we can obtain a milder version of Weyl's gravity, known
as conformal gravity. In this theory, we impose  
a local conformal invariance on the gravitational field action 
in the curved four-dimensional spacetime. 
The Einstein-Hilbert Lagrangian density for the
gravitational field is chosen on the requirement that the theory
of gravity is a second-order derivative theory. In conformal gravity,
the Lagrangian density is chosen on the principle of 
local conformal symmetry which is uniquely satisfied 
by the action $I_W = -\alpha\int d^4x\sqrt{-g}
C_{\mu\nu\kappa\lambda}C^{\mu\nu\kappa\lambda}$, where $C_{\mu\nu\kappa\lambda}$ is
the Weyl tensor, $\alpha$ is a coupling constant, and $g$ is the determinant
of the metric tensor $g_{\mu\nu}$. Conformal symmetry 
is garanteed by the invariance of the Weyl tensor to local conformal transformations
$g_{\mu\nu}(x) \to \Omega^2(x)g_{\mu\nu}(x)$, where $\Omega^2(x)$ is an arbitrary conformal
factor that can be specified by a spontaneous breaking of the conformal invariance
\citep{edery06, mannh08b}.
The theory of gravity implied by the action $I_W$ is a fourth-order derivative theory.
The vacuum exterior solution for a static and spherically
symmetric spacetime contains the Schwarzschild solution \citep{MK89}. 
The weak-field limit is consistent with the Solar System observational
data \citep{mannh07}, unlike claimed in previous investigations 
\citep{barabash99, flanagan06, barabash07}. 

\cite{mannh93, mannh97} shows that conformal gravity can reproduce
the rotation curves of disc galaxies without dark matter. Moreover, conformal 
gravity can explain the current accelerated expansion of the universe
without resorting to a fine-tuned cosmological constant or to the existence of dark energy
\citep{mannh01, mannh03, varieschi08}.
Unlike the standard theory, where the universe starts accelerating at redshift $z\la 1$,
conformal gravity predicts that the universe is accelerating
at all times. Therefore, observational
data probing the Hubble plot at very high redshift (e.g., \citealt{navia08, wei08}) can 
be a decisive test.

More recently, conformal theory 
has been proposed as a valid candidate for building a theory of quantum gravity \citep{mannh08};
in fact, theories based on fourth-order derivative equations of motion 
have had the long-lasting problem of suffering from the presence of ghosts. 
 \citet{bender08} have recently shown that this erroneous belief is the result
of considering the canonical conjugates ${\bf p}$ 
of the generic dynamical variables ${\bf q}$, when the ${\bf q}$'s are real,
to be Hermitian operators; 
however, this assumption is incorrect, and when the non-Hermiticity property of
the Hamiltonian of these higher-order quantum field theories is correctly
taken into account, the states with negative norm disappear.  

From an astrophysical perspective, however, the form of conformal gravity
that has been proposed in the literature currently has two main shortcomings: 
the abundance of light elements, and the gravitational lensing phenomenology.

Conformal gravity nicely avoids the requirement of an initial
Big Bang singularity, but it still predicts an early universe sufficiently dense
and hot to ignite the light element nucleosynthesis. 
However, conformal gravity predicts a too slow initial expansion rate.
This rate favours the destruction of most of the
deuterium produced in the early universe \citep{knox93, elizondo94} 
and poses conformal gravity in serious difficulties compared to 
the standard Big Bang nucleosynthesis. Conformal gravity 
necessarily requires astrophysical processes for the production of the
deuterium currently observed, for
example neutron radiative capture
on protons in the atmospheres of active stars \citep{mullan99} or
gamma-ray bursts \citep{inoue03}. However, the processes investigated to date 
do not seem to be efficient
enough. For example, significant production of deuterium in the Galaxy
seems to be ruled out \citep{proda03}.
  
The second open issue is the conformal gravity prediction of
gravitational lensing. Early investigations of gravitational lensing in conformal gravity 
\citep{walker94, edery98, edery99p, edery99}
show that, in the weak-field limit, the
deflection angle due to a point mass $M$ is $\Delta\alpha=4GM/c^2r - \gamma r$,
where $r$ is the radius of the photon closest approach; $\Delta\alpha$
contains the additional term $\gamma r$ when compared to the
General Relativity result. The constant $\gamma$ has to be positive to fit the
galaxy rotation curves, thus implying a repulsive effect in gravitational
lensing. It was later realized \citep{edery01} that the geodesics of
photons are independent of the conformal factor $\Omega^2(x)$ and one
can choose an appropriate conformal factor and a radial coordinate transformation to
yield attractive geodesics for both massive and massless particles.
However, in the strong-field limit, the light deflection might still be divergent or even impossible
\citep{pireaux04a, pireaux04b}.

Until these open questions are completely settled, conformal gravity 
cannot yet be considered ruled out
by observations. Moreover, conformal invariance plays a crucial
role in elementary particle physics and a viable theory of gravity that includes
this property can at least be suggestive of a relevant route towards the
unification of the fundamental interactions.

From the astrophysical point of view, conformal gravity can be further tested
by investigating the formation of cosmic structure. To date, 
nobody has yet explored how the large-scale structure forms in conformal gravity. 
If structures
form by gravitational instability, as in the standard theory, 
the development of a cosmological perturbation theory, which is 
still lacking, becomes inevitable. 
This theory would enable the comparison of conformal gravity 
with the spectrum of the Cosmic Microwave Background anisotropies
and would provide initial conditions for the simulation of the structure evolution into
the non-linear regime.

Before building up such a theory, however, it is useful to check whether conformal gravity 
is able to reproduce
the equilibrium configuration of cosmic structures, other than galaxies,
without dark matter.
\cite{horne06} has already shown that, if we interpret 
the observational data of the intracluster medium of X-ray clusters 
by assuming hydrostatic equilibrium, conformal gravity
requires a factor of ten less baryonic mass than inferred
from the X-ray surface brightness measures. 
Here, we extend this analysis by performing hydrodynamical simulations 
of self-gravitating gas clouds. 
We modify a standard Tree+SPH code to perform numerical simulations 
of self-gravitating systems in conformal gravity. 
The numerical tool we create is extremely relevant
if we eventually wish to investigate the formation of the large-scale structure,
because we will massively need to resort to numerical integrations 
when the evolution of the density perturbations
reaches the non-linear regime.

In Section \ref{sec:CG}, we review the basic steps leading to the 
gravitational potential of a static point source in conformal gravity. 
In Section \ref{sec:sphere}, inspired by the analysis of \citet{horne06}, 
we compute the gravitational potential energy of a spherical system,
and in Section \ref{sec:xray} we use this result to compute
the expected mean temperature and the temperature profile of the intracluster medium.
In Section \ref{sec:numerics}, we derive the same results with  hydrodynamical
simulations of self-gravitating gas clouds.

\section{Conformal Gravity}
\label{sec:CG}

Conformal gravity is a theory of gravity based on the action $I_W = -\alpha\int d^4x\sqrt{-g}
C_{\mu\nu\kappa\lambda}C^{\mu\nu\kappa\lambda}$. 
For any finite, positive, non-vanishing, continuous, real
function $\Omega^2(x)$ of the four spacetime coordinates $x$, this theory is
invariant for any conformal transformation $g_{\mu\nu}(x) \to \Omega^2(x)g_{\mu\nu}(x)$, because
the Weyl tensor is invariant for these transformations. The variational 
principle applied to the action leads to fourth-order field equations, whose
solution for the vacuum exterior to a static point source leads to the line element
\begin{eqnarray}
ds^2 & =& \Omega^2(x) \left\{ \left[1+{2\phi(r)\over c^2}\right] c^2 dt^2\right. \cr
& - & \left.\left[1+{2\phi(r)\over c^2}\right]^{-1} dr^2  
 - r^2 (d\theta^2 + \sin^2\theta d\phi^2)\right\}
\end{eqnarray}     
where $c$ is the speed of light,
\begin{equation}
{\phi(r)\over c^2} = -{\beta\over 2} {(2-3\beta \gamma)\over r}  - {3\over 2} \beta\gamma 
+ {\gamma\over 2} r  - {k\over 2} r^2 \; , 
\label{eq:metric}
\end{equation}
and $\beta$, $\gamma$ and $k$ are three integration constants. 
The arbitrary function $\Omega^2(x)$
can be specified by a mechanism which breaks the conformal symmetry. 
\citet{MK89} implicitly assume $\Omega^2(x)=1$. 

According to this assumption, when $\gamma=k=0$, the
metric (\ref{eq:metric}) reduces to the usual Schwarzschild metric with $\beta = GM/c^2$,
$M$ the gravitational mass of the point source and $G$ the gravitational constant. 
The term $kr^2/2$ corresponds to a cosmological solution which is conformal
to a Robertson-Walker background; therefore, $k$ can be chosen small enough
that it can be neglected on the scales of galaxies and galaxy clusters, and we will
not consider this term hereafter.
 Moreover, the rotation velocities of spiral galaxies 
suggest a value of $\gamma$ sufficiently small
that the terms proportional 
to $\beta\gamma\propto \gamma/c^2$ can be safely ignored \citep{mannh93}.
Equation (\ref{eq:metric}) thus reduces to  
\begin{equation}
{\phi(r)\over c^2} = -{\beta\over r} +{\gamma \over 2} r \; . 
\label{eq:potential}
\end{equation}
\citet{mannh97} 
uses the observed rotation curves of eleven spiral galaxies of widely different luminosities 
to constrain $\gamma$. It turns out that $\gamma$ also depends on the point source mass $M$. 
\citet{mannh97} suggests the parametrization $\gamma=\gamma_0 + M\gamma_*$, where
$\gamma_0=3.06\times 10^{-28}$~m$^{-1}$ and $\gamma_*=5.42\times 10^{-39}$~m$^{-1}$~M$_\odot^{-1}$;
$\gamma_0$ and $\gamma_*$ should be two universal physical constants.
In the following, we adopt a more convenient form of the potential
(\ref{eq:potential}) suggested by \citet{horne06}:
\begin{equation}
\phi(r)  =  - {GM\over r} + {GM\over R_0^2}r + {GM_0\over R_0^2}r  
\label{eq:phi-point-orig}
\end{equation}
where $M_0=(\gamma_0/\gamma_*) M_\odot = 5.6\times 10^{10}M_\odot$ and 
$R_0=(2GM_\odot/\gamma_*c^2)^{1/2}= 24$~kpc.

\section{Gravitational potential energy of a spherical system}
\label{sec:sphere}

\begin{figure*}
\centering
\includegraphics[angle=0,scale=.45]{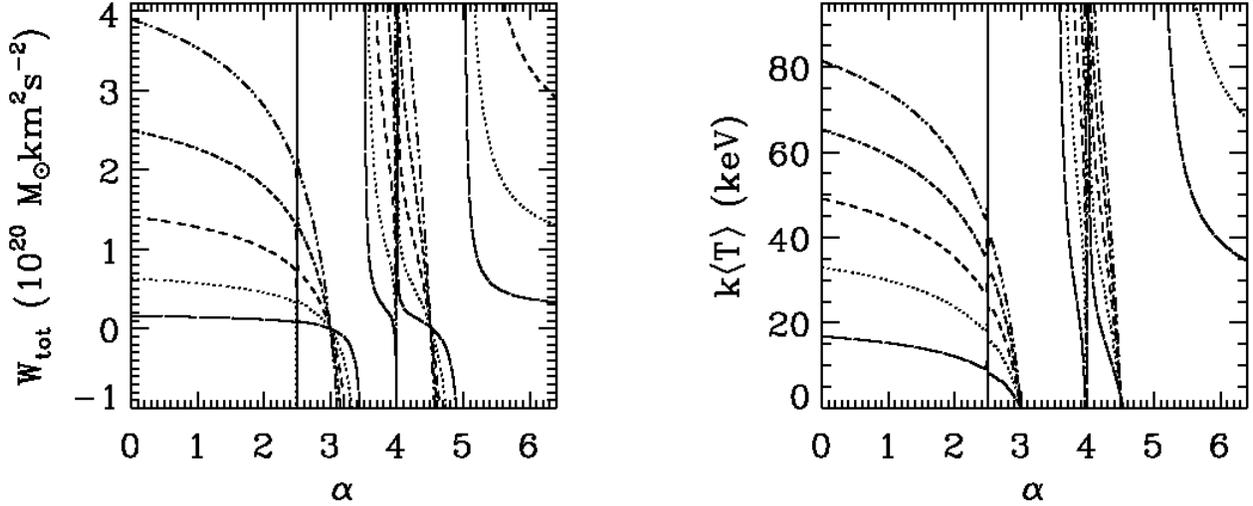}
\caption{ Left panel: Total gravitational potential energy
of a spherical system with a power-law density profile $\rho \propto r^{-\alpha}$, radius $a=1$ Mpc, and mass $M=2\times 10^{12}$~M$_\odot$ (solid line), $4\times 10^{12}$~M$_\odot$ (dotted line),
$6\times 10^{12}$~M$_\odot$ (dashed line), $8\times 10^{12}$~M$_\odot$ (dot-dashed line), 
$10^{13}$~M$_\odot$ (dot-dot-dot-dashed line); see Section \ref{sec:sphere} for details.
Right panel: The mean gas temperature of the systems
shown in the left panel according to the virial theorem; see Section \ref{sec:xray} for details.}
\label{fig:powerlaw-sphere}
\end{figure*}

The gravitational potential of a point source (equation \ref{eq:phi-point-orig})
\begin{equation}
\phi_{ps}(r) =  \phi_N(r)+\phi_{CM}(r) + \phi_{CC}(r) \; ,
\label{eq:phi-point}
\end{equation}
with $\phi_N(r) = -GM/r$ the usual Newtonian potential, $\phi_{CM}(r) = GMr/R_0^2$, and $\phi_{CC}(r
) = GM_0r/R_0^2$, generates the acceleration 
\begin{equation}
g_{ps}(r) = -\nabla\phi_{ps}= - {GM\over r^2} - {GM\over R_0^2} - {GM_0\over R_0^2} \; . 
\end{equation}

The last term is a constant acceleration independent of the mass
of the source of the gravitational field. The physical origin of this term is very
subtle. Unlike in Newtonian gravity, the presence of the linear term $\phi_{CM}(r)$ 
clearly prevents us from neglecting the contribution of distant objects 
to the acceleration of any body in the universe.  
According to \citet{mannh97, mannh06}, 
in conformal gravity the net effect of all the mass in the universe 
is to contribute a constant acceleration in addition to the acceleration
due to a local source. \citet{mannh97} shows that, with an appropriate
coordinate transformation allowed by the conformal invariance, the
mass in the universe exactly generates the additional linear potential $\phi_{CC}(r)$, 
that we name the {\it curvature} potential.

This argument implies that, to compute the gravitational potential energy of an extended source, 
we need to treat the {\it curvature} potential $\phi_{CC}(r)$ and the potential 
$\phi_N(r)+\phi_{CM}(r)$, originated by the local field source, separately. 
For the time being, let us consider this latter gravitational potential alone.

Consider an extended source. The gravitational potential energy of two of its mass elements 
with separation ${\bf r}$ is $dW_{AB}=dm_Adm_Bf({\bf r})$, where $f({\bf r})$ is the part
of the potential $\phi({\bf r})$ that depends on ${\bf r}$ only. When we sum
over all the mass elements, we get the total gravitational potential energy $W=(1/2)\int dm_A\int dm_Bf({\bf r})$, 
namely 
\begin{equation}
W = {1\over 2} \int \phi({\bf r})\rho({\bf r})d^3{\bf r} \; ,
\label{eq:pot-pot}
\end{equation}
where $\rho({\bf r})$ is the mass density distribution of the source and $\phi({\bf r})
= \phi_N({\bf r})+\phi_{CM}({\bf r})$ is
the gravitational potential generated by its self-gravity alone.

The total potential energy $W$ can also be derived with a different argument
which provides an alternative expression to equation (\ref{eq:pot-pot}).
 For a system of $N$ particles with position
${\bf r}_i$ and mass $m_i$, each feeling a force ${\bf F}({\bf r}_i)$, we can define 
the virial $\sum_i {\bf r}_i\cdot {\bf F}({\bf r}_i)$; the force 
is ${\bf F}({\bf r}_i) = -m_i\sum_j \nabla_{r_{ij}} \phi({\bf r}_{ij})$ where
${\bf r}_{ij}={\bf r}_i-{\bf r}_j$. Now, $m_i\nabla_{r_{ij}} \phi({\bf r}_{ij}) = - m_j 
\nabla_{r_{ji}} \phi({\bf r}_{ji})$
for Newton's third law, and we can write 
\begin{eqnarray}
 \sum_i {\bf r}_i\cdot {\bf F}({\bf r}_i) &  = & - 
\sum_i m_i\sum_{j<i}{\bf r}_i\cdot \nabla_{r_{ij}} \phi({\bf r}_{ij}) + \cr
& & \phantom{\sum_i } + \sum_i m_i\sum_{j>i} {\bf r}_j\cdot \nabla_{r_{ij}} \phi({\bf r}_{ij}) \cr
& = & -\sum_i m_i\sum_{j<i} {\bf r}_{ij}\cdot \nabla_{r_{ij}} 
\phi({\bf r}_{ij}) \; ,
\end{eqnarray}
since $\nabla_{r_{ij}} \phi({\bf r}_{ij})=0$ when $j=i$. 

Homogeneous functions of order $\lambda$ of
$M$ variables $({\bf x}_1,\cdots,{\bf x}_M)$ 
are defined by the relation 
\begin{equation} 
f(\alpha{\bf x}_1,\cdots,\alpha{\bf x}_M)= \alpha^\lambda f({\bf x}_1,\cdots,{\bf x}_M)
\end{equation}
for any non-null $\alpha$; they satisfy Euler's theorem 
\begin{equation}
\sum_i {\bf x}_i \cdot \nabla_{x_i}f({\bf x}_1,\cdots,{\bf x}_M) = \lambda f({\bf x}_1,
\dots,{\bf x}_M) \; .
\end{equation} 
In the potential $\phi=\phi_N+\phi_{CM}$, $\phi_N$ and $\phi_{CM}$ are homogeneous
functions of order $\lambda_N=-1$ and $\lambda_{CM}=1$ respectively. By applying Euler's theorem 
for $M=1$,
we can thus write 
\begin{eqnarray}
\sum_i {\bf r}_i\cdot {\bf F}({\bf r}_i) & = & -\sum_i \sum_{j<i} m_i[-\phi_N({\bf r}_{ij})
+ \phi_{CM}({\bf r}_{ij})]\cr & = & W_N-W_{CM} \; .
\label{eq:virial}
\end{eqnarray} 
Now, $W=W_N+W_{CM}$ and we obtain $W=2W_N-\sum_i  {\bf r}_i\cdot {\bf F}({\bf r}_i)$.
In the continuous limit 
\begin{equation}
W=2W_N+\int \rho({\bf r}) {\bf r}\cdot \nabla\phi({\bf r}) d^3 {\bf r} \; .   
\label{eq:pot-grad}
\end{equation}

To compute the gravitational potential $\phi(r)$ 
of a self-gravitating spherical system, we follow \citet{horne06} 
and we first consider a 
homogeneous spherical shell of density $\rho$, radius $R$ and mass $m=4\pi \rho R^2 {\rm d}R$.
At a generic point in space of coordinate ${\bf r}$, each mass element $\delta m
= \rho R^2 \sin\theta {\rm d}R {\rm d}\theta {\rm d}\varphi$ of the shell generates
the potential $\delta\phi(r)$ given by equation (\ref{eq:phi-point}). 
By setting the coordinate system such that ${\bf r}=(0,0,r)$ without
loss of generality, the element $\delta m$ has coordinates $R(\cos\varphi\sin\theta,\sin\varphi\sin\theta,\cos\theta)$
and generates the potential at ${\bf r}$
\begin{equation}
\delta \phi({\bf r}) = G\rho R^2\sin\theta{\rm d}R{\rm d}\theta{\rm d}\varphi \left(-{1\over x} + {x\over R_0^2}\right)\; ,
\end{equation}
where $x^2 = R^2 + r^2 -2rR\cos\theta$. By integrating over ${\rm d}\theta{\rm d}\varphi$, we
find the potential of the shell
\begin{equation}
\phi_{sh}(r) = Gm \left\{\begin{array}{ll}
-{1\over R} + {1\over R_0^2}\left({r^2\over 3R} + R\right) & r<R\\
-{1\over r} + {1\over R_0^2}\left({R^2\over 3r} + r\right) & r>R \; . \\
\end{array}\right.
\end{equation}

The gravitational potential of a self-gravitating sphere is thus
\begin{eqnarray}
{\phi(r)\over G}=-{I_0(r)\over r} - E_{-1}(r)& +& {1\over R_0^2}\left[{I_2(r)\over 3r} 
+ rI_0(r)\right.\cr 
 & + & \left. {r^2\over 3}E_{-1}(r) + E_1(r)\right] 
\label{eq:phi-ext}
\end{eqnarray} 
where $I_n(r)=4\pi \int_0^r \rho(x)x^{n+2} dx$ and $E_n(r)=4\pi \int_r^{+\infty} \rho(x)x^{n+2}dx$,
and the acceleration is
\begin{equation}
-{\nabla\phi(r)\over G} = -{I_0(r)\over r^2} + {1\over R_0^2}\left[{I_2(r)\over 3 r^2} - {2\over 3} r E_{-1}(r) 
- I_0(r)\right] \; .
\end{equation}

The total gravitational potential $W$ of a sphere can now be computed with either equation 
(\ref{eq:pot-pot}) or equation (\ref{eq:pot-grad}).
For a power-law density profile $\rho(r)=\rho_0(r/a)^{-\alpha}$ of a
system truncated at radius $a$ with total mass $M=4\pi\rho_0 a^3/(3-\alpha)$, 
with $\alpha\ne 3$, we find $I_0(r)=M (r/a)^{3-\alpha}$,
$I_2(r)=[(3-\alpha)/(5-\alpha)] M a^2(r/a)^{5-\alpha}$,
$E_{-1}(r) = [(3-\alpha)/(2-\alpha)](M/a)
[1-(r/a)^{2-\alpha}]$, and $E_1(r)=Ma(3-\alpha)/(4-\alpha)[1-(r/a)^{4-\alpha}]$,
and the gravitational potential energy is 
\begin{eqnarray}
W & =& W_N\left[1-\left(a\over R_0\right)^2 {2(5-2\alpha)(9-2\alpha)\over 3(7-2\alpha)(5-\alpha)} \right] \cr
& \equiv& W_N[1-h(\alpha)]\; ;
\label{eq:gravpot}
\end{eqnarray}
here
\begin{equation}
W_N = -{GM^2\over a}{3-\alpha\over 5-2\alpha} \;  
\end{equation}
is the potential energy in Newtonian gravity, that is recovered in the limit $R_0\to\infty$. 

To compute the contribution $W_{\rm curv}$ of the curvature acceleration $-GM_0/R_0^2$ to be 
added to $W$, we can use either equation (\ref{eq:pot-grad}) or equation (\ref{eq:pot-pot})
without the factor $(1/2)$, because the {\it curvature} mass $M_0$ is {\it not} accelerated by
the mass $\rho({\bf r})d^3{\bf r}$. We find
\begin{equation}
W_{\rm curv} = GMM_0{a\over R_0^2}{3-\alpha\over 4-\alpha}=-W_N\left(a\over R_0\right)^2
{M_0\over M} {5-2\alpha\over 4-\alpha}\; .
\end{equation}

 The left panel of Figure \ref{fig:powerlaw-sphere} shows the total gravitational potential
energy 
\begin{equation}
W_{\rm tot}=W_N(\alpha)[1-h(\alpha)]+W_{\rm curv}(\alpha)
\label{eq:gravpottot}
\end{equation}
of a spherical system as a function of $\alpha$ and its total mass $M$. 
When $\alpha\to 2.5, 3.5, 4$ and $5$, $W_{\rm tot}\to\pm\infty$.
In Figure \ref{fig:powerlaw-sphere}, 
the pole of $W_N$ at $\alpha=2.5$ seems 
to have different properties than the other poles, but it is only a graphic artifact. In fact, 
the contribution of conformal gravity to $W_{\rm tot}$ 
is proportional to $(a/R_0)^2\sim 10^3$, and it always 
dominates the newtonian $W_N$: we must have $\vert \alpha-2.5\vert \ll 10^{-3}$
to see $W_{\rm tot}\to\pm\infty$. 

Finally, the ratio between the curvature potential and the conformal gravity potential due
to the sphere self-gravity is $\sim M_0/M$; therefore, 
$W_{\rm curv}$ is negligible when $M\gg M_0$, as in our case, where $M_0/M\sim 10^{-2}$,
unless, of course, $\alpha\to 4$.

\section{X-ray clusters}\label{sec:xray}

\begin{figure*}
\includegraphics[angle=0]{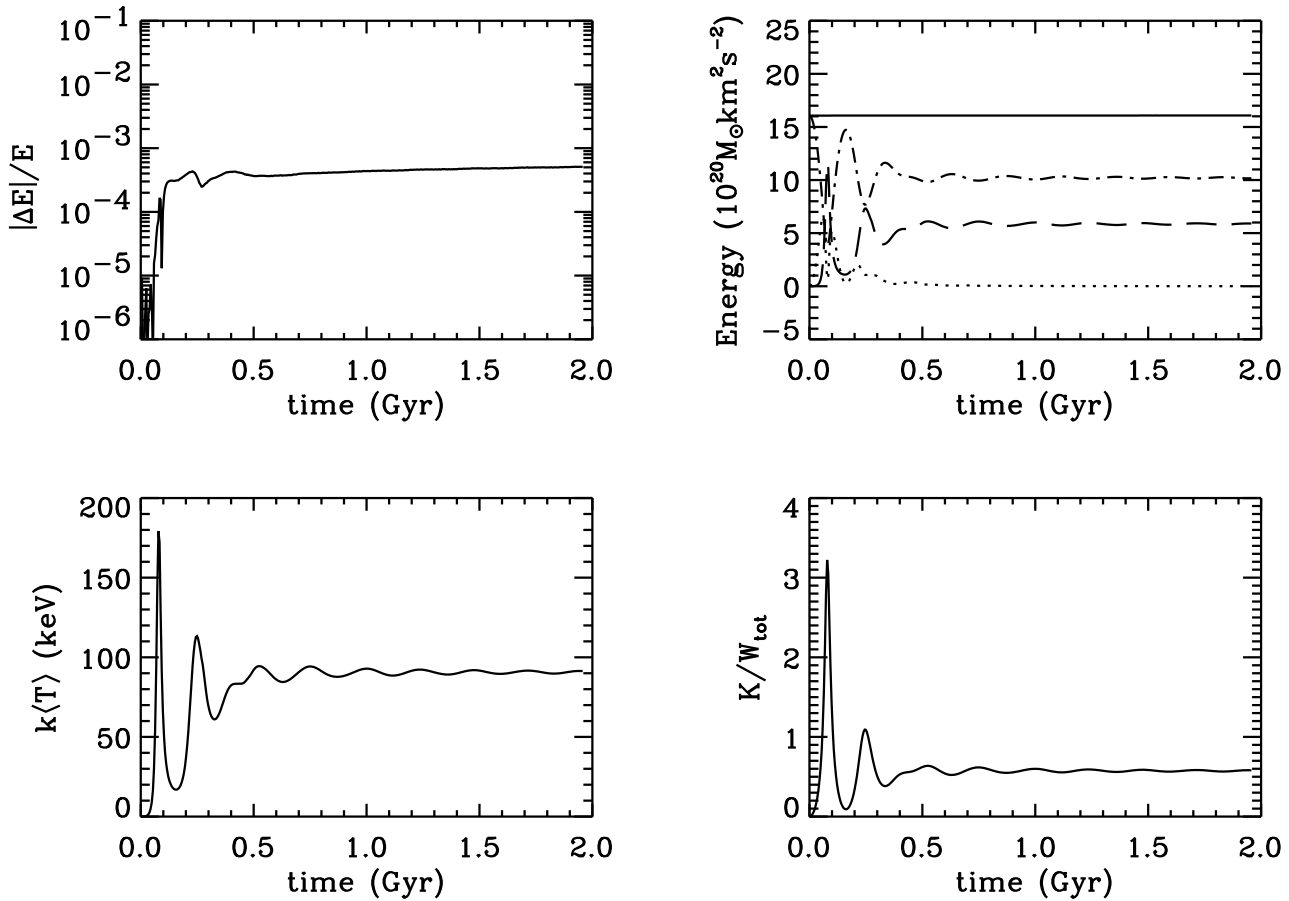}
\caption{Upper left panel: Fluctuations of the total energy. Upper right panel: Evolution of the
kinetic (dotted line), thermal (dashed line), gravitational potential (dot-dashed line), 
and total (solid line)
energy. Lower left panel: Evolution of the mass-weighted mean X-ray temperature. Lower right panel: 
Evolution of the virial ratio. }
\label{fig:energy-4pan}
\end{figure*}

A system of $N$ particles of 
position ${\bf r}_i$ experiencing the force ${\bf F}_i$ is in virial equilibrium when it satisfies 
the virial theorem relation $2K + \sum {\bf r}_i\cdot {\bf F}_i=0$, where $K$ is the total kinetic
energy of the system.  We saw above (equation \ref{eq:virial}) 
that $\sum {\bf r}_i\cdot {\bf F}_i = 2W_N-W_{\rm tot}$ in conformal
gravity. Note that we replaced $W_{CM}$ with $W_{\rm tot}$ to include the curvature potential. 

Observations indicate that in X-ray clusters the 
total mass of galaxies is roughly ten times smaller than the total mass 
of the intracluster medium (e.g., \citealt{white93}). 
 We can therefore model X-ray clusters
as individual clouds of hot gas with no galaxies and, of course,
with no dark matter. This approximation is partly inappropriate for
cD clusters, where the contribution of the cD galaxy to the gravitational
potential in the cluster center is relevant; however, cD clusters are $\sim 20\%$
of the population of nearby clusters \citep{rood71} and our analysis should 
be valid for most clusters.

If the gas bulk flows and 
turbulent motions are negligible, the total kinetic
energy is $K = (3/2)(\gamma-1)U = (3/2)M k\langle T\rangle/\mu m_p$, where $U=[M/(\gamma-1)]k\langle T\rangle/\mu m_p$ 
is the internal energy of the gas of density $\rho({\bf r})$, $M$ is the gas total mass,
$\langle T\rangle=\int\rho({\bf r})T({\bf r})d^3{\bf r}/\int\rho({\bf r})d^3{\bf r}$ 
is its mass-weighted mean temperature, 
$k$ the Boltzmann constant, $m_p$ the proton mass, $\mu$ the mean molecular
weight, and $\gamma$ the adiabatic index. The virial theorem\footnote{The 
virial theorem can also be generalized in $f(R)$ gravities 
\citep{bohmer08}.} thus yields 
\begin{equation}
k\langle T\rangle = -{ \mu m_p\over 3M} (2W_N-W_{\rm tot}) 
\label{eq:virial-th}
\end{equation}
Figure \ref{fig:powerlaw-sphere}
shows the mean temperatures $k\langle T\rangle$ 
of spherical systems with a power-law density profile 
as a function of $\alpha$ and its total mass $M$.
 
Consider an isothermal sphere with $\alpha=2$, radius $a=1$ Mpc, and mass $M=10^{13} M_\odot$.
Newtonian gravity predicts $k\langle T\rangle=0.09$ keV. Conformal gravity
predicts $k\langle T\rangle=58.21$ keV, namely a temperature $(a/R_0)^2\sim 10^3$ larger. 
To explain a more typical temperature of, say, $k\langle T\rangle\sim 6$ keV 
without resorting to  dark matter, the gas mass should be $M\sim 10^{12} M_\odot$, a factor
of ten lower than the mass of the intracluster medium measured from the 
X-ray surface brightness. 

This result agrees with the claim of \citet{horne06} that the acceleration 
provided by the conformal gravity potential is too intense. We now show
that the disagreement with observations extends to the temperature
profile, besides its normalization.

To compute the temperature profile of a self-gravitating spherical system in equilibrium, 
we need to solve the Boltzmann equation coupled to the Poisson equation.
In conformal gravity, the Poisson equation $\nabla^2\phi = 4\pi G\rho$
is replaced by a fourth order equation $\nabla^4\phi \propto \rho$ \citep{MK94}. 
Following this approach is not a trivial task, either
analytically or numerically. In the next section, we resort
to a smoothed-particle-hydrodynamical simulation to obtain a self-consistent solution to
this problem.

In this section, to illustrate qualitatively the expected result, we  
can consider the first moment of the Boltzmann equation, namely the Euler
equation, which, in hydrostatic equilibrium, reduces to the relation
$\nabla p=-\rho\nabla\phi$, where $p$ is the gas pressure. 
Provided that the density profile $\rho(r)$ is known,
for an ideal gas with equation of state $p=\rho kT/\mu m_p$,
this equation immediately integrates to 
\begin{equation}
{kT(r)\over \mu m_p} = -{1\over \rho(r)}\left[\int \rho(r) \nabla \phi(r) dr + A_0\right]\;  
\label{eq:hydro-temp}
\end{equation}
where $A_0$ is an integration constant. At large radii, real clusters usually have the intracluster
medium density
profile which decreases more rapidly than the temperature profile. This constraint requires
that the term $A_0/\rho(r)$ should disappear at large radii and implies $A_0=0$.
For a power-law density profile, we find
\begin{eqnarray}
T(r) & =& T_N(r)   
\left\{1 + {\alpha-1\over (2-\alpha)^2 } \left(a\over R_0\right)^2\left[{2
\over (5-\alpha)}\left(r\over a\right)^2 -\right.\right.\cr
& & \left.\left.{4(3-\alpha)\over 3} \left(r\over a\right)^\alpha\right]\right\}
+ {GM_0\over R_0} {1\over 1-\alpha}\left(r\over R_0\right)
\label{eq:T-profile}
\end{eqnarray}
where
\begin{equation}
{kT_N(r)\over \mu m_p} = {GM\over 2a} {1\over \alpha-1}\left(r\over a\right)^{2-\alpha} \; .
\end{equation}
Note that for an isothermal sphere ($\alpha=2$) in Newtonian gravity, 
this equation yields $kT_N/\mu m_p= GM/2a$.
This temperature disagrees with the correct virial theorem result 
$k T_N/\mu m_p=GM/3a$ (equation \ref{eq:virial-th}).  
This discrepancy is a simple consequence of the inconsistent approach
we used to obtain equation (\ref{eq:T-profile}).
However, we can use equation (\ref{eq:T-profile}) to predict that, at large
radii, the temperature profile increases at least as $r^2$.

\begin{figure}
\centering
\includegraphics[width=0.9\hsize,angle=0]{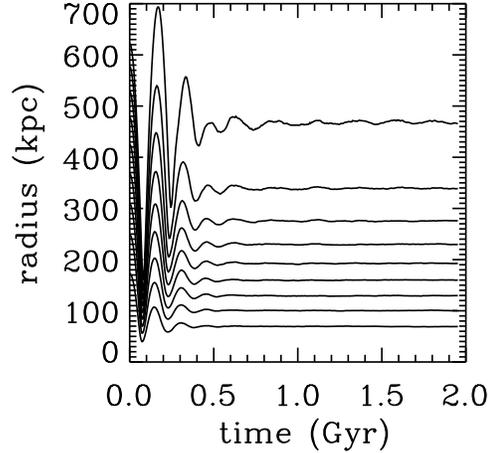}
\caption{Evolution of the size of the gas cloud: From bottom to top, lines are for radii containing
10\% to 90\% of the total cloud mass.}
\label{fig:radius}
\end{figure}

We thus conclude that the values of the conformal
physical constants $R_0=24$~kpc and $M_0=5.6\times 10^{10}$~M$_\odot$ 
implied by the galaxy rotation 
curves provide X-ray clusters with average temperatures a factor of 
ten too large, and rising 
temperature profiles, clearly at odds with real clusters.

\section{Numerical solution}
\label{sec:numerics}

To solve properly the coupled Boltzmann and Poisson equations and 
predict the correct X-ray properties of a galaxy cluster in conformal gravity, we
run hydrodynamical simulations of isolated spheres of gas with a 
Smoothed-Particle-Hydrodynamic (SPH) code. We use a modified version of the 
publicly available GADGET-1.1 code \citep{springel01,springel05}. 
The original version of GADGET-1.1 is
a Tree+SPH code which integrates the equations of motion of
$N$ particles which interact gravitationally. The $N$ particles
are a discrete representation of either a collisionless or a collisional fluid; 
traditional simulations 
contain both collisionless and collisional particles which represent the dark matter
and the baryonic fluids, respectively. 

In our simulations, we only require the presence of the baryonic
fluid and we only have collisional particles. 
Moreover, in Newtonian
gravity, the potential between two particles decreases with 
the interparticle distance, and GADGET-1.1 arranges the particles in
a tree structure in order to average the contribution of
distant particles to the local acceleration. This technique
substantially decreases the time required to compute 
the acceleration of individual particles. Unfortunately, in conformal gravity this
technique can not be applied, because the gravitational
potential contains a term that increases linearly with
the interparticle distance, and the contribution
of distant particles to the gravitational potential
must be computed individually. We will therefore use GADGET-1.1
as a direct $N$-body integrator. The computational cost thus 
increases with $N^2$ rather than $N\ln N$.
In appendix \ref{app:SPH-kernel}, we describe in detail
the modification we introduced in GADGET-1.1 to 
compute the interparticle forces in conformal gravity.

We simulate the evolution of isolated spherical clouds of gas 
with vacuum boundary conditions. 
We neglect radiative cooling. 
 At the end of this section we discuss how simulating
the gas physics more realistically can actually be relevant
to find a way to alleviate the discrepancy between conformal gravity and 
the observed thermal properties of X-ray clusters.

A gas cloud is represented by $N=4096$ gas particles. 
The chosen $N$ is substantially smaller than state-of-the-art
simulations of galaxy clusters and is due
to the $N^2$ scaling of the computational cost. We do not
actually need a larger $N$, because the aim of our simulations is to test 
the general viability of conformal
gravity rather than the detailed physics of the intracluster medium, which would require
higher spatial and mass resolutions. In fact,
the current form of the conformal gravitational potential yields clusters so different
from real clusters that running costly simulations is unwarranted.

\begin{figure*}
\centering
\includegraphics[width=0.45\hsize,angle=0]{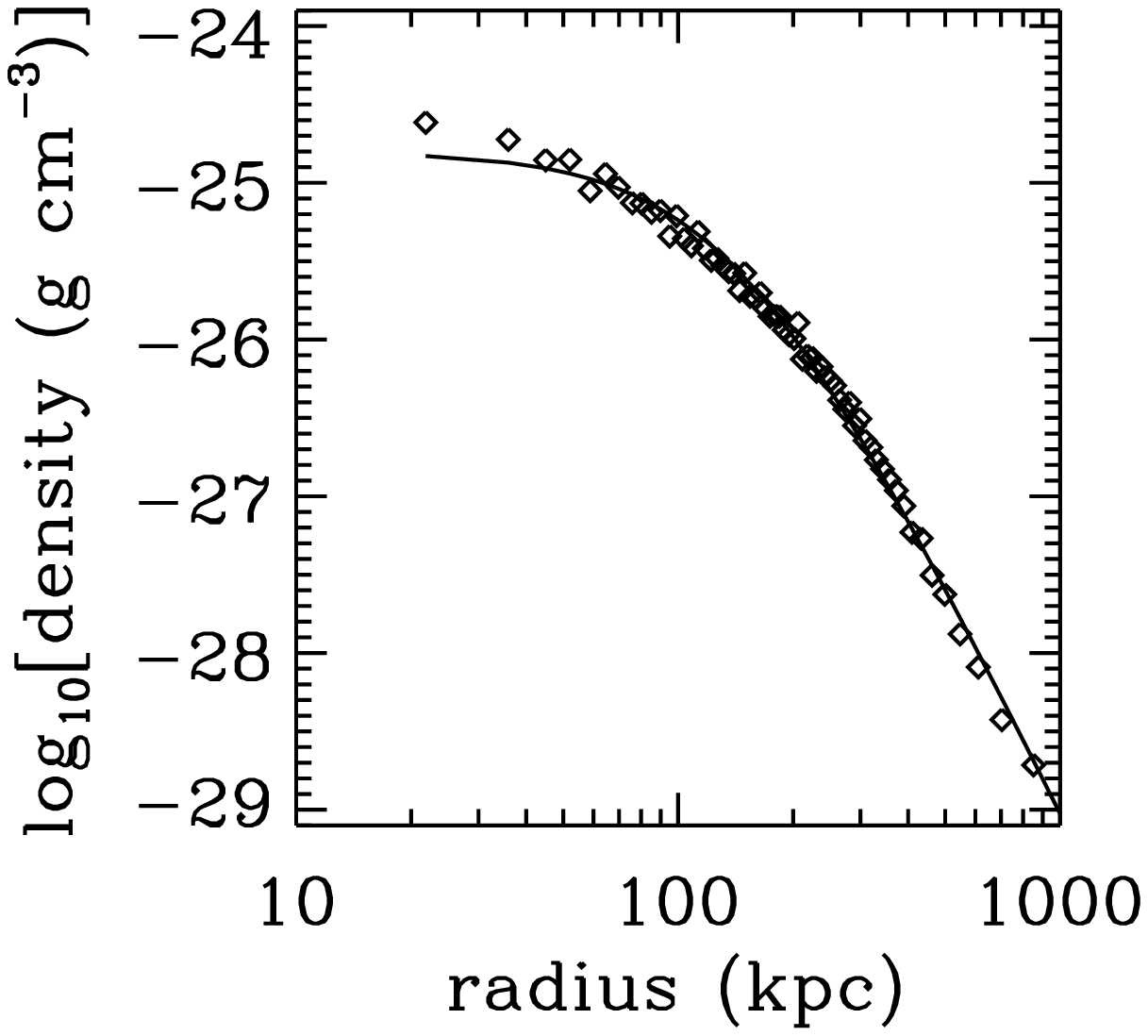}
\includegraphics[width=0.45\hsize,angle=0]{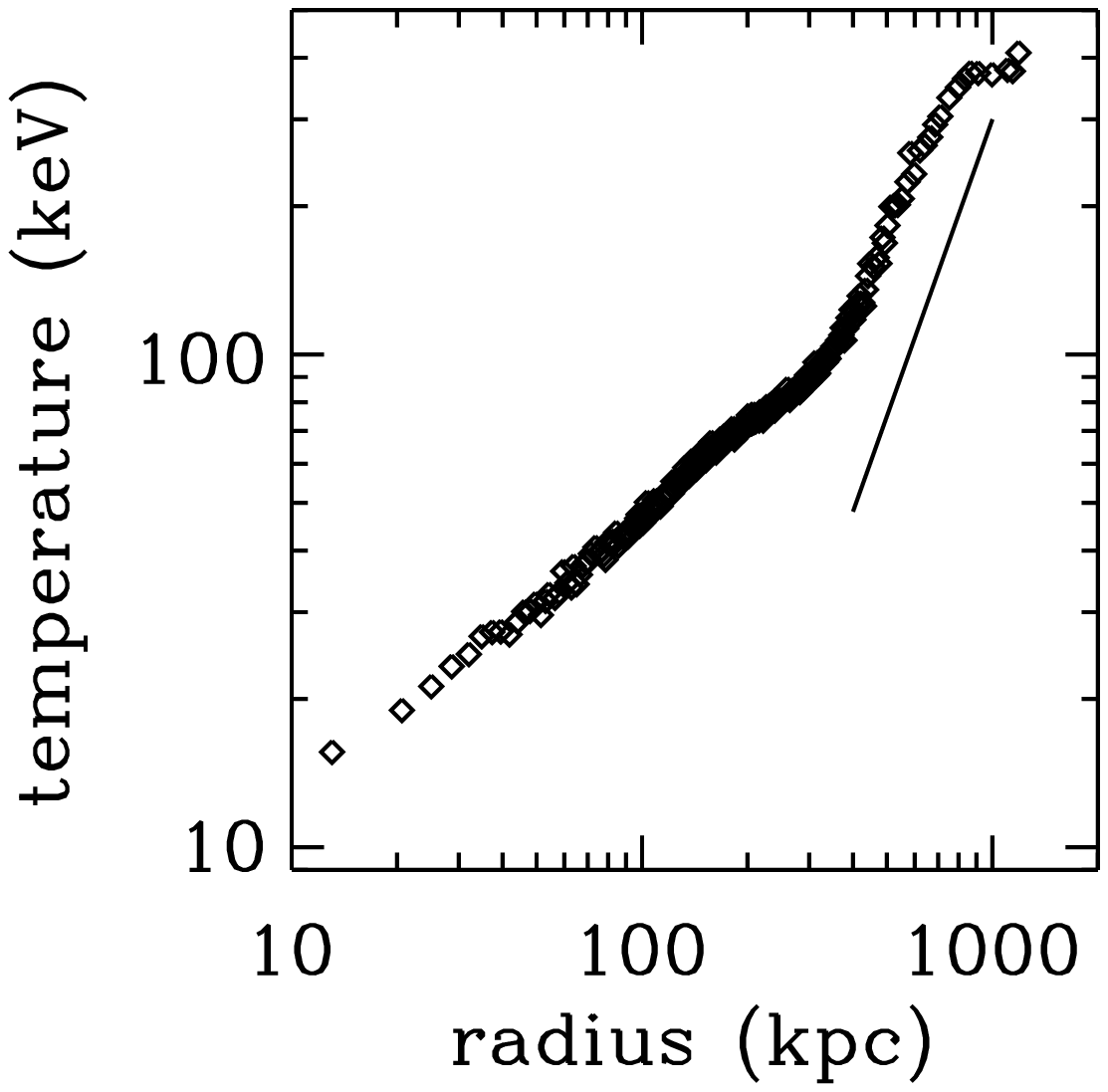}
\caption{Mass density (left panel) and temperature (right panel) profiles of
the final configuration of the gas cloud. The solid line in the left panel is the best-fit 
$\beta$-model profile. The straight line in the right panel shows the
$T\propto r^2$ slope.}
\label{fig:profiles}
\end{figure*}

\begin{figure*}
\centering
\includegraphics[width=0.45\hsize,angle=0]{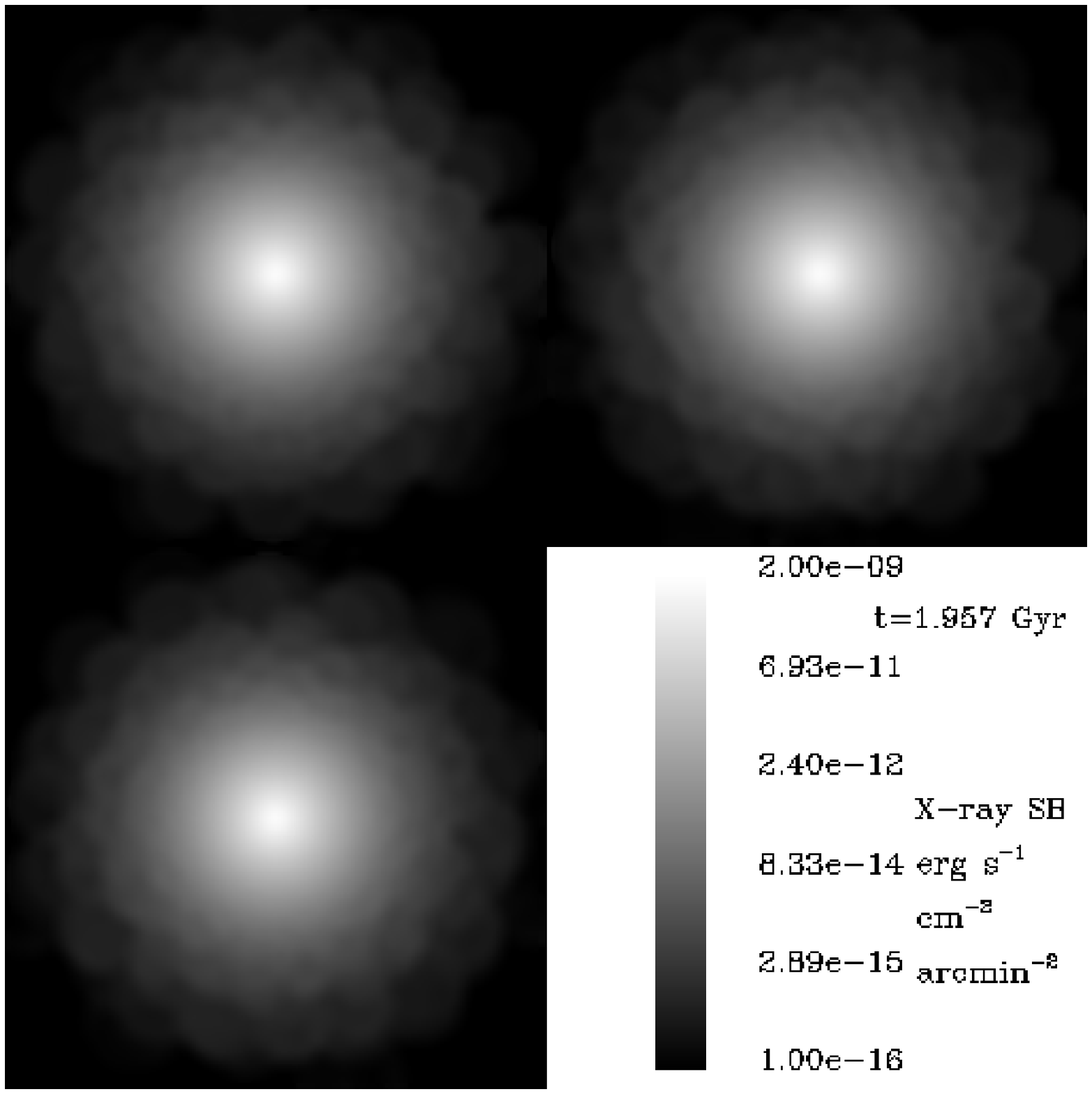}
\includegraphics[width=0.45\hsize,angle=0]{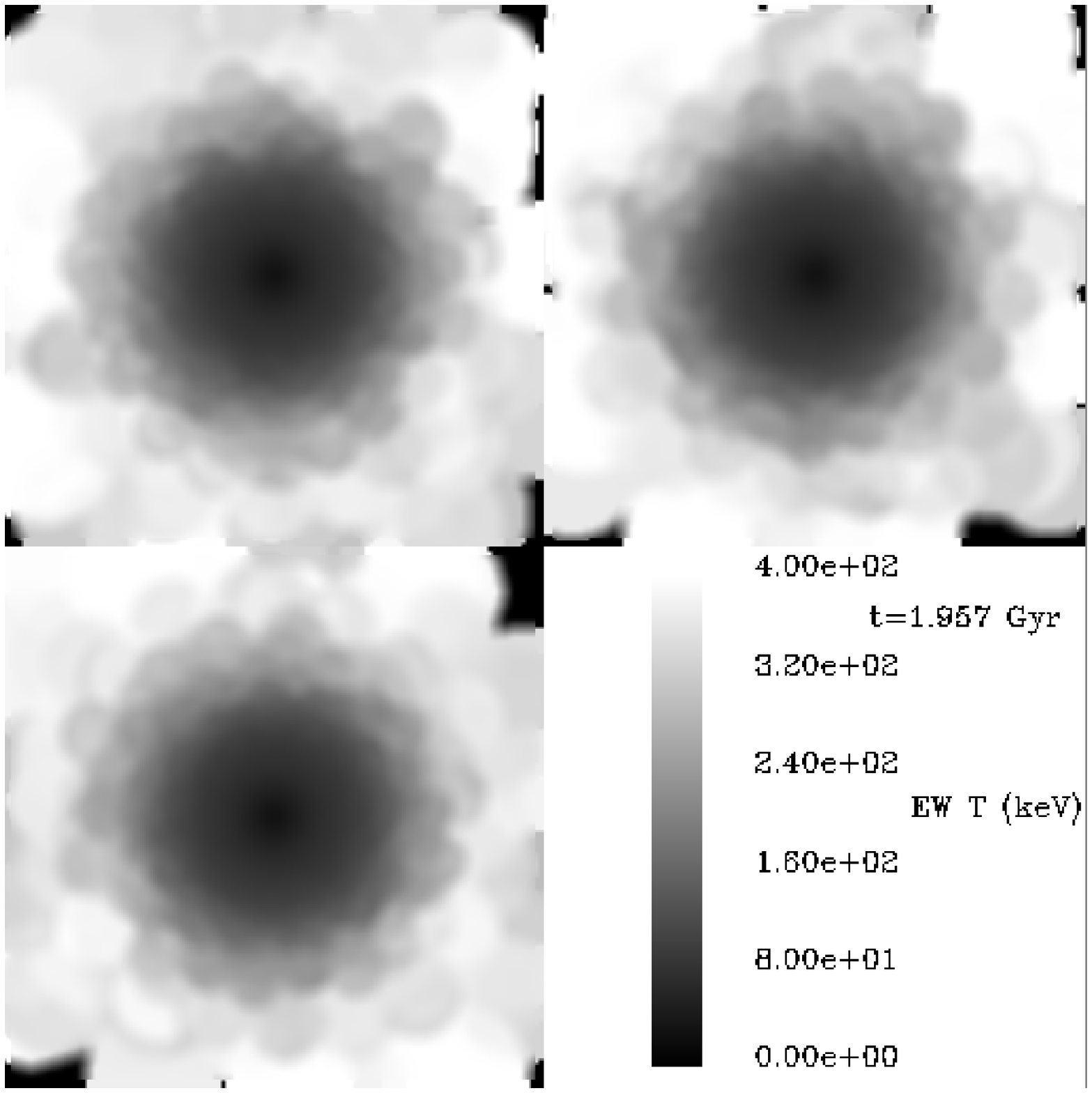}
\caption{X-ray surface brightness (SB) (left panel) and emission weighted (EW) temperature 
(right panel) maps of
the final configuration of the gas cloud. Each panel shows three orthogonal projections. 
Each map is 2~Mpc on a side.}
\label{fig:maps}
\end{figure*}

The gas cloud initially has a $\beta$-model density profile
\begin{equation}
\rho(r)=\rho_0\left[1+\left(r\over r_0\right)^2\right]^{-3\beta/2}\; 
\end{equation}
and null thermal and kinetic energies. Each gas particle has
gravitational softening length $\epsilon=0.1$~kpc. 
Each gas particle also has an individual SPH softening parameter $h$ used
to estimate the thermodynamical quantities, as mentioned below; the
sphere with radius $h$ centered on the particle contains a fixed
number $N_b$ of neighbor particles. We set $N_b=40$ in our simulations. 

In the following, we show the result of a typical simulation: 
a cloud of total mass $M=2.7\times 10^{13}$~M$_\odot$, and
initial values of the $\beta$-model $\beta=0.636$, $r_0=134$~kpc, and
$\rho_0=1.52\times 10^{-26}$~g~cm$^{-3}$. 
The upper left panel of Figure \ref{fig:energy-4pan} shows 
that the total energy is conserved to better then 0.1\%. The evolution
of the gravitational potential, kinetic, thermal and total energies (upper right panel) 
shows that in half a billion years the system reaches equilibrium
and remains confined; in fact, 90\% of the mass is
within $500$~kpc (Figure \ref{fig:radius}). 
It follows that in conformal gravity
a cloud of gas can easily remain confined without the need of dark matter.

The bottom right panel of Figure \ref{fig:energy-4pan} shows 
the evolution of the virial ratio: the sum $K$ of the thermal and
kinetic energy over the total gravitational potential energy $W_{\rm tot}$. 
At equilibrium we must have 
$2K+ \sum {\bf r}_i\cdot {\bf F}_i =0$, where $\sum {\bf r}_i\cdot {\bf F}_i = 2W_N-W_{\rm tot}$
(Section \ref{sec:xray}). Therefore $K/W_{\rm tot}=(1/2) - (W_N/W_{\rm tot})$. 
From quation \ref{eq:gravpottot} we see that $W_{\rm tot}$ 
is positive and $W_{\rm tot}\gg \vert W_N\vert$ 
because $(a/R_0)^2\gg 1$; 
therefore, unlike Newtonian gravity where $K/W_{\rm tot}=-1/2$ at equilibrium, here 
the virial ratio $K/ W_{\rm tot}$ is slightly larger than $1/2$. 

In the Tree+SPH code each gas particles of mass $m_i$ and density $\rho_i$
carries the electron number density ${n_e}_i$,  and
the internal energy $u_i$. By assuming an optically thin gas
of primordial cosmic abundance $X=0.76$ and $Y=0.24$, we can 
estimate the particle temperature $T_i$; we also estimate the ion number density
${n_{\rm H}}_i + {n_{\rm He}}_i$ that we use below. 
The bottom left panel of Figure \ref{fig:energy-4pan} shows
that the final mass-weighted mean temperature of the system is $90$~keV, as expected from
the argument provided in Section \ref{sec:xray}.

Figure \ref{fig:profiles} shows the density and temperature
profiles of the final configuration. 
As in real clusters, the $\beta-$model is still an excellent approximation to the final
density profile. The best fit yields $\beta=1.65$, $r_0=142$~kpc and 
$\rho_0=1.6\times 10^{-25}$~g~cm$^{-3}$. These parameters roughly agree with
typical values of real clusters, although $\beta$ is $\sim 2-3$ times larger 
than typical values (see, e.g., \citealt{mark99}).

The result in strong disagreement with observations is the rising temperature profile shown 
in the right panel of Figure \ref{fig:profiles}. At large radii,
we see that $T\propto r^2$, as anticipated in Section \ref{sec:xray}. 

For completeness, we show the surface brightness and the
temperature maps in Figure \ref{fig:maps}.
Each map is an equally spaced $N_p\times N_p$ grid, with $N_p=128$, 
corresponding to a length resolution $\approx 15.6$~kpc.
In the Tree+SPH code, each
gas particle has a smoothing length $h_i$ and the thermodynamical
quantities it carries are distributed within the sphere of radius
$h_i$ according to the compact kernel $W(r; h_i)$, which 
has the same functional form of the gravitational kernel (see Appendix \ref{app:SPH-kernel}),
where $r$ is the distance to the particle center.
The X-ray surface brightness $S_{jk}$ on the grid point $\{j,k\}$ is
\begin{equation}
S_{jk} = {1\over d_p^2} \sum {n_e}_i ({n_{\rm H}}_i + {n_{\rm He}}_i) \Lambda(T_i) w_i {\rm d}V_i
\end{equation}
where $d_p^2$ is the pixel area, the sum runs over all the
particles, and $w_i \propto\int W(x){\rm d}l$ is the weight proportional to
the fraction of the particle volume ${\rm d}V_i=m_i/\rho_i$ which
contributes to the grid point $\{j,k\}$. For each particle, the
weights $w_k$ are normalized to satisfy the relation $\sum w_k=1$
where the sum is now over the grid points within the particle
circle. When $h_i$ is so small that the circle contains no grid point,
the particle quantity is fully assigned to the closest grid point.
We use the cooling function $\Lambda(T)$ of \cite{suth93}. 
The total X-ray luminosity is 
\begin{equation}
L_X = \sum {n_e}_i ({n_{\rm H}}_i + {n_{\rm He}}_i) \Lambda(T_i) {\rm d}V_i\; .
\end{equation} 
Given the large equilibrium temperature, the cluster shown
in Figure \ref{fig:maps} has $L_X=3.86\times 10^{46}$~erg~s$^{-1}$ at equilibrium,
an order of magnitude larger than the most luminous X-ray clusters.
The temperature whose map is shown in Figure \ref{fig:maps} is
the emission-weighted temperature
\begin{equation}
T_{jk} = {\sum {n_e}_i ({n_{\rm H}}_i + {n_{\rm He}}_i) \Lambda(T_i) T_i w_i {\rm d}V_i
	\over \sum {n_e}_i ({n_{\rm H}}_i + {n_{\rm He}}_i) \Lambda(T_i) w_i {\rm d}V_i}\; .
\end{equation}
This map dramatically shows how the thermal properties of the intracluster medium totally
disagree with real clusters.

 At this point, we cannot draw our conclusions without the following relevant cautionary tale.
The results we show here 
derive from simulations that treat the gas adiabatically. 
In other words, we neglect radiative cooling and the gas heating
processes due to astrophysical sources, as supernovae
explosions, energy injection from active galactic nuclei, or
galactic and stellar winds. These heating processes act at
different times and with different effectiveness, depending on
the detailed history of star and
galaxy formation. This history is currently totally unexplored in conformal gravity,
and the detailed mechanisms of the heating processes themselves are still poorly understood
(e.g., \citealt{borgani08}).
Therefore, the appropriate inclusion of these cooling and heating processes 
in our simulations is beyond the
illustrative purpose of this paper. 

However, an appropriate interplay between cooling and heating
processes could in principle provide a possible route to reconcile 
conformal gravity with the properties of X-ray clusters.
In fact, since our results show that, for the typical gas density of
real systems, the gas temperature 
is an order of magnitude larger than observed,
radiative cooling would be very efficient in conformal clusters 
and the entire intracluster medium
could cool in less than 1~Gyr. Effective heating processes would thus be
necessary to reheat the intracluster medium to the observed
X-ray temperatures. These heating processes must be more efficient
in the cluster center than in the cluster outskirts, because
the cooling time increases with the clustrocentric radius.
However, although we cannot exclude that
the various complex physical processes could eventually conspire to reconcile
conformal gravity with observations, it
is plausible that, in order to provide intracluster media with
the observed thermal properties, the cooling and heating processes
should be severely fine-tuned during the formation and 
evolution of clusters and cluster galaxies.

\section{Conclusion}

Conformal gravity can explain the rotation curves of disk galaxies
and the current accelerated expansion of the universe without
resorting to dark matter and dark energy.  
We have modified a Tree+SPH code to run hydrodynamical simulations
of isolated X-ray galaxy clusters to show that conformal gravity
does not share the same success on the scales of clusters.

These simulations confirm our simple analytic estimates 
that show that gas clouds with mass $\sim 10^{12}-10^{13}$~M$_\odot$,
which are typical values of the total mass of the hot gas present in real clusters, remain
confined with an equilibrium mean temperature $\sim 10-100$~keV,
ten times larger than the observed temperatures; more dramatically, 
because of the presence of a linear term in the gravitational
potential, at large clustrocentric radius $r$, the gas temperature increases with $r^2$, rather
than decreasing as in real systems.

 Our analysis totally neglects radiative cooling and gas heating from
astrophysical sources, as supernovae or active galactic nuclei. 
The interplay between these processes can in principle provide a way to reconcile conformal
gravity with observations. It is however unclear if and how much these processes
should be fine-tuned to provide X-ray clusters in agreement with observations. 

In addition to this topic, we can see two more open 
issues whose solution might also reconcile conformal
gravity with observations:

\begin{enumerate}
\item In conformal gravity, all the matter in the universe is expected to affect the local
dynamics. The net effect is to contribute a constant inward acceleration $-GM_0/R_0$ 
in addition to the gravitational acceleration generated by
local sources. Because we included this constant acceleration in our simulations, 
we could neglect the rest of the universe and impose 
vacuum boundary conditions. It might be possible that assimilating the 
gravitational influence of the nearby matter surrounding the X-ray cluster  
in the constant ``universe'' acceleration $-GM_0/R_0$ is inappropriate:
in fact, nearby external matter might decrease the gravitational
attraction of the interior matter and hopefully reduce the thermal energy of the gas.
To appropriately investigate this effect, 
we should simulate the dynamics of large-scale structure within a full
cosmological context. However, this task is not trivial just because the
gravitational field is highly non-local.
This investigation would also benefit from the implementation 
into the numerical simulation of the yet unavailable
theory of structure formation.
\item The gravitationl potential we implemented in 
our simulations derives from 
a metric where the conformal invariance is broken by
an arbitrary choice of the conformal factor $\Omega^2(x)$. 
It is unclear whether this choice provides a coordinate
system whose physics describes the real world or it is
an artifact of the reference frame.
It also remains to be seen whether a spontaneous breaking of the 
conformal invariance, in theories where matter
and gravity are conformally coupled \citep{edery06}, can provide a metric, and thus
a gravitational potential, where the observed thermal
properties of the intracluster medium can be reproduced. 
\end{enumerate}

In Section \ref{sec:intro} we mentioned that the nucleosynthesis
of light elements and the phenomenology of gravitational lensing
are two open issues that need to be solved before accepting
conformal gravity as a viable
alternative theory of gravity and cosmology.
Here, we have shown that the thermodynamics of X-ray clusters
poses a third challenge to this theory.

\section*{ACKNOWLEDGMENTS}
We thank Volker Springel for making public his superb Tree+SPH code
and the referee whose careful report 
prompted us to clarify some aspects of our results. We 
acknowledge vivid discussions with J\"org Colberg, Keith Horne 
and Philip Mannheim on very early versions of this work.
Support from the PRIN2006 grant ``Co\-stituenti fondamentali dell'Universo'' of
the Italian Ministry of University and Scientific Research and from the
INFN grant PD51 is gratefully acknowledged.

\appendix

\section{The softening kernel }
\label{app:SPH-kernel}

The publicly available code GADGET-1.1 \citep{springel01,springel05} 
is a Tree+SPH code which integrates the equations of motion
of $N$ particles which interact gravitationally. The particles
are a discrete representation of either a collisionless or a collisional fluid. 
Here, we consider only adiabatic processes and neglect the
possibility of radiative cooling of the collisional fluid.

The only modification to the code we need is the 
computation of the gravitational potential $\phi$ and its corresponding
acceleration. At position ${\bf r}$, $N$ particles of mass $m_i$
at position ${\bf x}_i$ generates the potential 
\begin{equation}
\phi({\bf r})  = G\sum_{i=1}^N \left[- m_i g_N(y_i) + {m_i\over R_0^2}  g_{CM}(y_i) + 
{M_0\over R_0^2} g_{CC}(y_i)\right]
\end{equation} 
where $y_i=\vert{\bf r}-{\bf x}_i\vert$. For point sources, $g_N(y)=1/y$
and $g_{CM}(y)=g_{CC}(y)=y$.
The acceleration ${\bf a}({\bf r}) = -\nabla \phi({\bf r})$ is
\begin{equation}
{\bf a}({\bf r})  = G\sum_{i=1}^N {\bf y}_i \left[m_i g^1_N(y_i) - {m_i\over R_0^2}  g^1_{CM}(y_i) -
{M_0\over R_0^2} g^1_{CC}(y_i)\right]
\end{equation}
where $g^1_N(y)y = dg_N/dy$, and analogously for $g^1_{CM}$ and $g^1_{CC}$.

GADGET-1.1 treats particles as extended spherical objects with mass $m$ and
density profile $\rho(r) = mW(r; h)$ where
\begin{displaymath}
W(r; h)={8\over \pi h^3}\left\{\begin{array}{ll}
1 - 6x^2 + 6x^3  &  0\le x < {1\over2} \\
 2(1-x)^3       &         {1\over 2}\le x< 1 \\
  0            &       x\ge 1\; , \end{array} \right.
\end{displaymath}
where $x=r/h$. $W(r; h)$ is a spline kernel which avoids unrealistic divergences of the Newtonian
acceleration for arbitrary small particle separations.

The spline kernel implies that each particle is not a point source of gravitational 
potential. Rather, its gravitational potential is correctly computed as an extended source
of density $W(r; h)$. According to equation (\ref{eq:phi-ext}) 
we find, with $u=y/h$, 
\begin{displaymath}
g_N(y; h) = {1\over h} \left\{\begin{array}{ll}
{14\over 5} -{16\over 3} u^2 + {48\over 5} u^4 -{32\over 5} u^5 &  0\le u < {1\over2} \\
-{1\over 15 u} +{16\over 5} -{32\over 3} u^2 + 16u^3 \\
{\phantom{-{1\over 15 u}}} -{48\over 5}u^4 + {32\over 15} u^5 &         {1\over 2}\le u< 1 \\
{1\over u} &       u\ge 1\; ,
\end{array} \right.
\end{displaymath}

\begin{displaymath}
g_N^1(y; h) = {1\over h^3} \left\{\begin{array}{ll}
-{32\over 3} +{192\over 5} u^2 -32 u^3  &  0\le u < {1\over2} \\
{1\over 15 u^3} -{64\over 3} +48 u -{192\over 5} u^2 \\
	{\phantom{{1\over 15 u^3}}} + {32\over 3}u^3  &         {1\over 2}\le u< 1 \\
-{1\over u^3} &       u\ge 1\; ,
\end{array} \right.
\end{displaymath}

\begin{displaymath}
g_{CM}(y; h) =  h \left\{\begin{array}{ll}
{31\over 70} +{14\over 15} u^2 - {8\over 15} u^4 +{16\over 35} u^6 \\
{\phantom{31\over 70}} - {8\over 35} u^7 &  0\le u < {1\over2} \\
-{1\over 840 u} +{16\over 35} -{u\over 15} +{16\over 15}u^2   \\ 
 {\phantom{-{1\over u}}} -{16\over 15}u^4 + {16\over 15} u^5 \\
 {\phantom{-{1\over u}}} -{16\over 35}u^6 +{8\over 105}u^7 &         {1\over 2}\le u< 1 \\
u + {3\over 40 u} &       u\ge 1\; ,
\end{array} \right.
\end{displaymath}

\begin{displaymath}
g_{CM}^1(y; h) = {1\over h} \left\{\begin{array}{ll}
{28\over 15} -{32\over 15} u^2 +{96\over 35}u^4 -{8\over 5}u^5  &  0\le u < {1\over2} \\
{1\over 840 u^3} -{1\over 15u} +{32\over 15} -{64\over 15} u^2  \\
{\phantom{1\over u^3}} + {16\over 3}u^3  -{96\over 35}u^4 + {8\over 15}u^5  &         {1\over 2}\le u< 1 \\
{1\over u} -{3\over 40 u^3} &       u\ge  1\; ,
\end{array} \right.
\end{displaymath}
$g_{CC} = y$, and $g^1_{CC}=1/y$.

\end{document}